# Improving energy efficiency in MANET's for healthcare environments

Sohail Abid [1], Imran Shafi [2] and Shahid Abid [3]

[1,3] Foundation University Rawalpindi Campus, Pakistan.

rsohailabid@yahoo.com

[2] Department of Computing and Technology

Abasyn University, Islamabad Campus, Pakistan.

imran.shafi@abasyn.edu.pk

ABSTRACT. *Now a day ad hoc mobile networks (MANETs) have lots of routing protocols, but no one can meet maximum performance. Some are good in a small network; some are suitable in large networks, and some give better performance in location or global networks. Today, modern and innovative applications for health care environments based on a wireless network are being developed in the commercial sectors. The emerging wireless networks are rapidly becoming a fundamental part of every single field of life. Our proposed DEERP framework gives a better performance as compared to other routing protocol.*

1. **Introduction.** Today advance applications of health care environments have been developed for ad-hoc networks. The capability to enhance health care telemetry with wearable miniature wireless sensors would have a deep impact on several ways of medical practice. In terms of efficiency the small portable wireless devices plays an important role in the health care environment and to provide essential support to patients. Wireless health care monitors/ equipments are available in the market for example, blood pressure monitors [1, 2], pulse Oximeters [3, 4], maternal uterine and fetal heart rate monitors [5], Wireless ECG System [6], and EKGs (Electrocardiographs) [7, 8, 9]. During disaster recovery or a group of casualty, the doctor fixes small sensors on each patient and monitor the results using Laptop and PDAs. In fig 1 (a), (b) and (c) some wireless devices used in the health care environment.

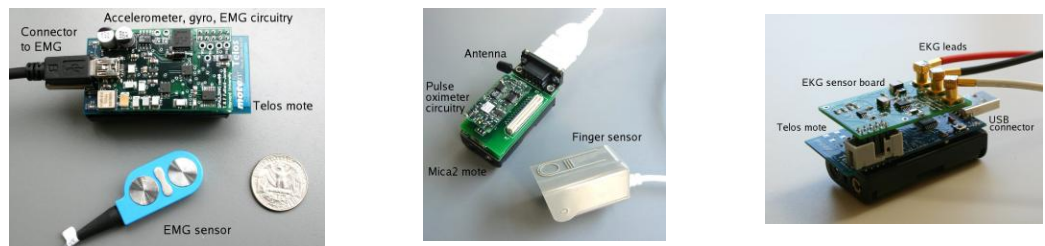

Fig 1:    (a) Motion capture and EMG        (b) Pulse Oxi-meter                     (c) EKG

In the health care environment, a huge importance is placed on data integrity, availability and security. Now a day private and public key security is implemented in the security of data in ad-hoc networks [10, 11, 12]. Due to mobility and congestion packet loss may be compromised.   Multiple doctors or nurses can receive patient data using multicast semantics support on the network layer. Due to mobility of patients, doctors and nurses quickly route changes, therefore energy efficient and multi-hop routing protocols are required in health care environments.

Ad hoc network building a network without any structure and set of hosts who are agreed to establish a connection or communication with each other exclusive of any centralized administration [13]. Each mobile device in an ad - hoc network acts as a router. Wireless MANET is ideal for every application of our daily life due to its mobility and accessibility everywhere. Now a day Mobile Ad hoc Network (MANET) is a quickly rising technology, due to its self-motivated topology and unique nature of scattered resources. Wireless MANET is rapidly emerging and trendy topology in almost all fields like medical, healthcare, banking and commerce, etc. Currently, wireless MANETs are becoming very popular and many routing protocols have been suggested by researchers. We give a brief introduction to wireless networks and discuss the issues and challenges regarding to performance and efficient use of energy. We are concerned with energy efficiency and select some well-known energy efficient routing protocols, and simulate these protocols in NS2 and analyze energy efficiency in different cases. MANET has principles due to which controls the number of hosts and route all packets between the mobile hosts in their wireless networks. MANET faces a lot of challenges, but we focus energy efficiency and performance. The energy efficiency and performance evaluation are two very important and critical challenges for routing protocols. Some routing protocols to have maximum packet delivery ratio and throughput, some have less end-to-end delay and minimum routing overhead; some consume less energy during idle mode; some consume less energy during receive mode, and some consumes less energy during transmit mode. On the other hand, some routing protocols to perform skillfully in small networks and some perform skillfully in large networks.

A mobile Ad-hoc network–MANET consists of mobile devices that are placed without any predefine pattern and regularly changing their position and connected with each other. The main aim of routing protocol is to discover routes from source to destination node and use the best and most efficient route. In case of route error the routing protocol switches to the other suitable route. During the establishment of and preservation of the route a less overhead and bandwidth utilization should be made [14]. We discuss in this research article, how healthcare networks use ICT (Information and Communications Technology) has been developed, and what sort of impact they have on the present health care system. We consider the fittings of mobile technologies in healthcare environments. We spotlight on WPAN (Wireless Personal Area Network) technologies, explicitly, IEEE 802.11 standard. Our research work has based upon the recent results on the energy efficiency and performance of routing protocols. Our proposed framework is to choose appropriate routing protocols, which give good performance the necessary implementation changes required to incorporate existing routing protocols in our framework. In our framework, we use only proactive and reactive routing protocols. The proposed framework presents wide assessment, using a very famous network simulator NS2. On the basis of our results that our proposed framework improves energy efficiency and performance as compared with other selected routing protocols.

2. **Related work.** It is specified in recent research [15] that each protocol is appropriate for a certain user environment and network. Some protocol that adapts the route construction method performs well in less mobility and decreases their performance in the high mobility environment. L. M. Feeney presented in his paper a comparison of energy consumption for DSR, AODV in NS2 [16]. Furthermore, in [17], [18], it proves that in the path creation method, competence is also a limiting factor; as the number of nodes increases the available throughput of each node approaches zero. Dr. S. P. Setty and B. Prasad compares QOS in energy consumption for proactive and reactive routing protocols with the impact of network size [19]. On the other hand, the support approach always succeeds in delivering messages [20]. Ved Prakash, Brajesh Kumar and A. K. Srivastava analyze and compare the energy efficiency of topology based, and location based routing protocols [21]. In this, review papers Neeraj Tantubay, Dinesh Ratan Gautam and Mukesh Kumar Dhariwal present a summary of different energy control techniques and various powers saving methods have been proposed in his research articles [22]. Feeney L. M. divides the methods which are used in energy-efficient awareness routing protocols in ad-hoc networks [23]. In first method when a host transmitting packets, the routing protocol minimized the total energy consumed during transmitting [24], [25], [26]. In a second method, load balance between hosts to increase the lifetime of the whole network, instead of managing energy consumption for individual packet [27], [28], [29]. A. Bamis and his group members present a mobility sensitive method [30]. Moreover, they find the importance of both technologies with respect to scalability issues. The pulse Oxi-meter equipped with 802.11 has developed by WiiSARD group [31] and EKG by SMART team [32]. Some researcher making monitoring infrastructures of daily activity of patient [33], monitoring infrastructures of daily activity of the patient at home [34], monitoring infrastructures of daily activity of patient at hospital [35].

## 3. Proposed DEERP Framework.

In our propose DEERP framework, we select two proactive and reactive routing protocols and used as one routing protocol. On the basis of previous research paper results [36], we combine the two best routing protocols and get better performance and energy efficiency. The structure of the proposed DEERP framework is shown in fig 2.

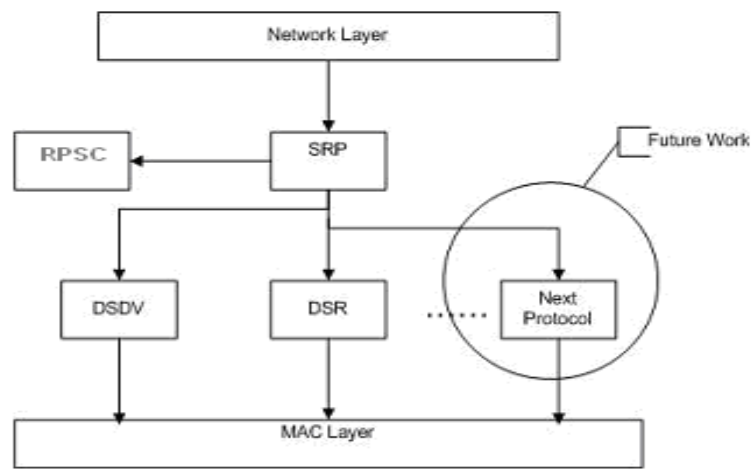

**Fig 2: Proposed DEEPR**

Propose a method, select one state like idle mode and check which protocol gives best performance in idle mode according to the given routing conditions. According to this method, we select routing protocols for above three modes. In other words, our proposed method selects different routing protocols based on their best performance and places these protocols in the above modes. In our proposed DEERP, a framework has one service. The function of this service is divided into three parts. In a first part SRP (Select Routing Protocol) service selects the best routing protocol for idle mode from RPSC (Routing Protocol Selection Criteria) table. In the second part, SRP services select routing protocol for TX mode, which is best in the RPSC table. In the third part, SRP serviced select routing protocol for Rx mode, which is best in the RPSC table. According to this process, we select two or three routing protocols. The combination of these protocols gives excellent performance.

The RPSC table consists of routing environment and characteristics like number of nodes, mobility model, node speed and best routing protocols on their performance with respect to RPSC table criteria. The SRP service checks the routing environment characteristics or conditions like number of nodes, mobility model, node speed, etc. and compares these characteristics with RPSC table. The protocol which performs well, according to the given routing conditions will be selected. In our case RPSC table is as below.

### 3.1. Routing Protocol Selection Criteria (RPSC)

The routing protocol is selected on following properties.

| Mobility Model | No. of Nodes | Node Speed | Mode | Best Performance Protocol |
|---|---|---|---|---|
| RWP | 5 - 25 | 1 – 10 ms | Idle | DSR |
| | | | Tx | DSDV |
| | | | Rx | DSR |
| | | | | |
| RPGM | 20- 80 | 0.5- 5 ms | Idle | DSDV |
| | | | Tx | DSDV |
| | | | Rx | DSR |
| | | | | |

Table 1: Routing Protocol Selection Criteria Table (RPSC)

### 4. Simulations

### 4.1. Methodology Used in Our Simulation-I
The network parameters used in our simulation is described in table 1.

| Simulation I Parameters | |
|---|---|
| Parameters | Values |
| MAC Type | IEEE 802.11 |
| Antenna | Omni directional |
| Simulation Time | 900 sec |
| Transmission range | 500 x 500 – 2000 x 2000 |
| Node speed | 0.5m/s to 5.0 m/s |
| Traffic Type | CBR |
| Data payload | 512 bytes/ packet |
| Packet rate | 8 packet/sec |
| Node Pause Time | 0 |
| Mobility Model | RPGM |
| Interface Queue Type | Drop Tail/Priori Queue |
| Interface Queue Length | 50 |
| No. of Nodes | 20 to 80 |

Table 1: Simulation I Parameters

### 4.2. Methodology Used in Our Simulation-II

The new changes in parameter used in our simulation II are described in table 2.

| Simulation II Parameters | |
|---|---|
| Parameters | Values |
| Simulation Time | 300 sec |
| Transmission range | 600 x 600 m |
| Mobility Model | Random Waypoint |
| No. of Nodes | 5, 10,15,20,25 |

Table 2: Simulation II Parameters

### 4.3. Energy Consumption Model:

There are four states of energy consumption of mobile devices which are given in table 3.

| Energy Consumption Parameters | |
|---|---|
| ei: | Energy Consumption during    Idle mode |
| es: | Energy Consumption during    Sleep mode |
| et: | Energy Consumed during Transmitting mode |
| er: | Energy Consumed during Receiving mode |

Table 3: Energy States

### 4.3.1. Transmit Mode (Tx)

Tx = (Pkt-size x 330) / 2 x $10^6$     And     PTx = Tx / TTx

Where PTx is transmitting power, Tx is transmitting energy and TTx is time take during packet transmit and Pkt-size is the size of packet in bits.

### 4.3.2. RX Mode

RX = (Pkt-size x 230) / 2 x 106 And     PRX = RX / TRX

Where PRX is receiving power, RX is receiving energy and TRX is time take during receiving a packet and Pkt-size is the size of packet in bits.

### 4.3.3. Idle/ Listening Mode

PIdle = PRX

Where PRX is power consumed in receiving mode and PIdle is power consumed in idle mode.

## 5. Results

Simulation I                                                Simulation II

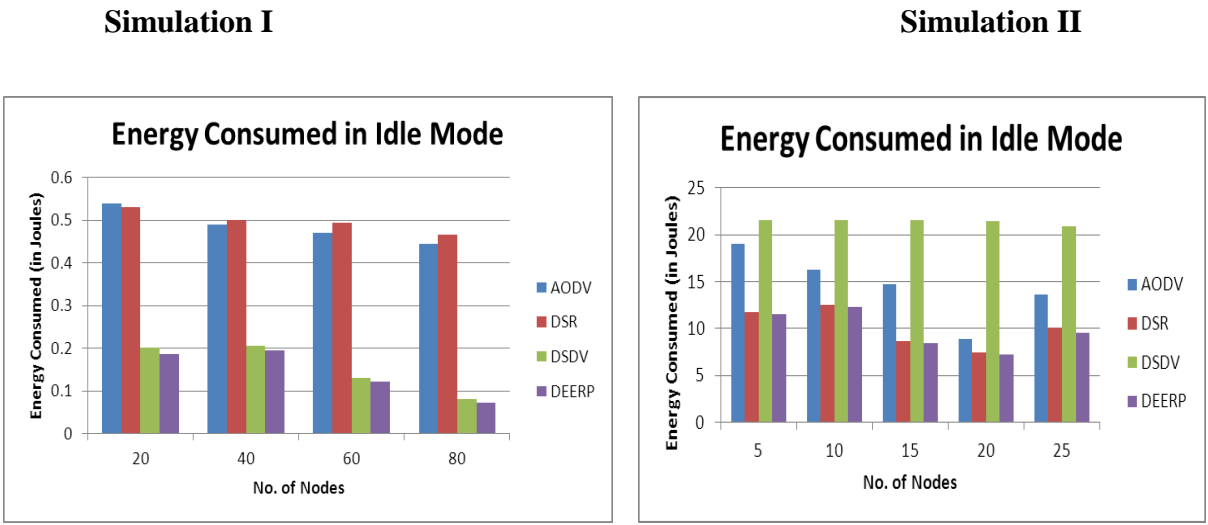

Fig 3: Avg. Energy consumed in idle mode.

In fig 3, it is clear that our proposed DEERP framework consume minimum energy in idle mode as compared to other protocols.

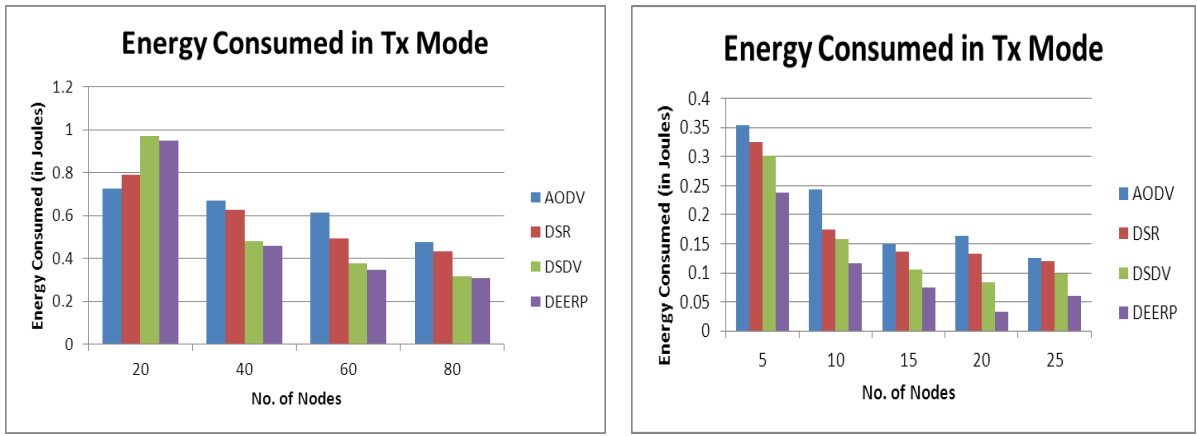

Fig 4: Avg. Energy consumed in Tx mode.

In fig 4, it is clear that our proposed DEERP framework consume minimum energy in Tx mode.

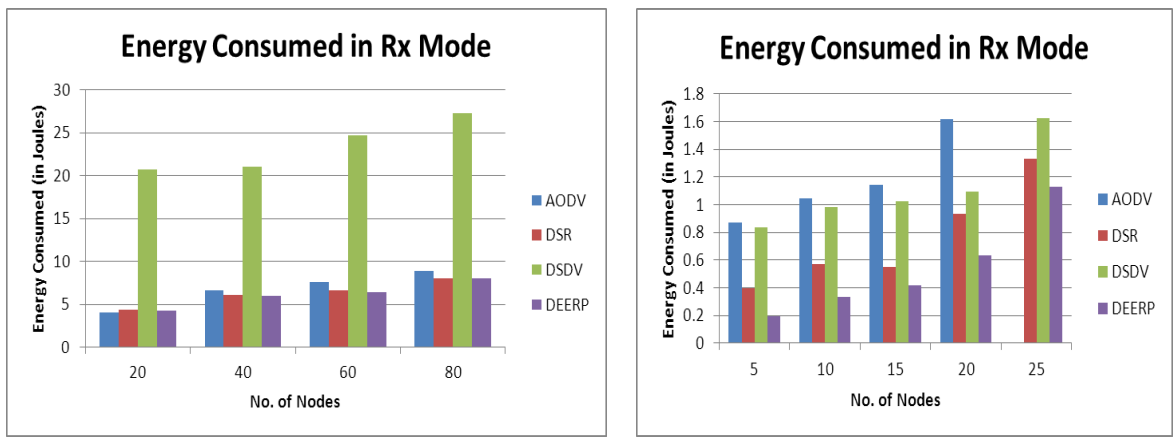

Fig 5: Avg. Energy consumed in Rx mode.

In fig 5, it is clear that our proposed DEERP framework consume minimum energy in Rx mode as compared to other protocols.

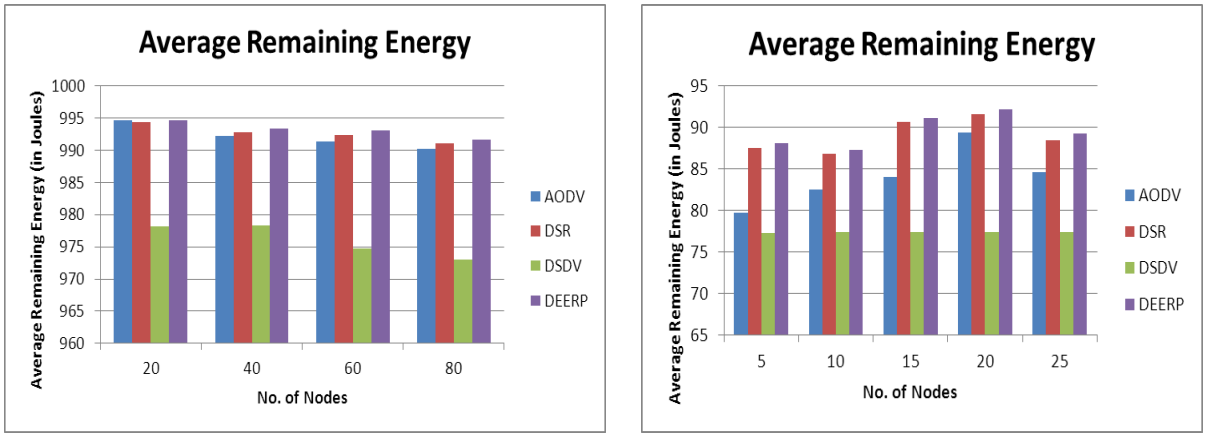

Fig 6: Avg. Remaining Energy

In fig 6, it is proved that our proposed DEERP framework has maximum remaining energy as compared to other protocols.

In the light of above graphs, it is proved that our proposed framework "DEERP" is a best routing framework in our scenario.

**Conclusion and Future Work**

In this research article, we discuss and compare four routing protocols to investigate the performance and energy consumption. In the light of above investigation we found that our proposed routing framework (DEERP) gives better performance as compared to other routing protocols. Our RPSC table selection criteria of routing protocol are limited because our selected routing protocols are selected from proactive and reactive routing protocols. In future work, we will select routing protocols from proactive, reactive, hybrid and location base protocols and improves RPSC table. We use two mobility models with limited criteria, In future it will be enhanced to improve the performance of the framework/protocol.